\documentclass{aastex62}
\usepackage[utf8]{inputenc}
\newcommand{\nh}{N$_{H}$}

\begin{document}

\title{The role of complex ionized absorber in the soft X-ray spectra of Intermediate Polars}

\author{Nazma Islam}
\affil{Center for Space Science and Technology, University of Maryland, Baltimore County, 1000 Hilltop Circle, Baltimore, MD 21250, USA}
\affil{X-ray Astrophysics Laboratory, NASA Goddard Space Flight Center, Greenbelt, MD 20771, USA}

\author{Koji Mukai}
\affil{Center for Space Science and Technology, University of Maryland, Baltimore County, 1000 Hilltop Circle, Baltimore, MD 21250, USA}
\affil{CRESST II and X-ray Astrophysics Laboratory, NASA Goddard Space Flight Center, Greenbelt, MD 20771, USA}

\correspondingauthor{Nazma Islam}
\email{nislam@umbc.edu}
\correspondingauthor{Koji Mukai}
\email{Koji.Mukai@umbc.edu}

\submitjournal{ApJ}

\begin{abstract}
In magnetic Cataclysmic Variables (mCVs), X-ray radiation originates from the shock heated multi-temperature plasma in the post-shock region near the white dwarf surface. These X-rays are modified by a complex distribution of absorbers in the pre-shock region. The presence of photo-ionized lines and warm absorber features in the soft X-ray spectra of these mCVs suggests that these absorbers are ionized. We developed the ionized complex absorber model {\tt zxipab}, which is represented by a power-law distribution of ionized absorbers in the pre-shock flow. Using the ionized absorber model {\tt zxipab} along with a cooling flow model and a reflection component, we model the broadband Chandra/HETG and NuSTAR spectra of two IPs: NY Lup and V1223 Sgr. We find that this model describes well many of the H and He like emission lines from medium Z elements, which arises from the collisionally excited plasma. However the model fails to account for some of the He like triplets from medium Z elements, which points towards its photo-ionization origin. We do not find a compelling evidence for a blackbody component to model the soft excess seen in the residuals of the Chandra/HETG spectra, which could be due to the uncertainties in estimation of the interstellar absorption of these sources using Chandra/HETG data and/or excess fluxes seen in some photo-ionized emission lines which are not accounted by the cooling flow model. We describe the implications of this model with respect to the geometry of the pre-shock region in these two IPs. 
\end{abstract}

\keywords{stars: white dwarfs, X-rays: NY Lup, X-rays: V1223 Sgr, methods: data analysis}

\section{Introduction}
Cataclysmic Variables (CVs) are interacting binaries consisting of a white dwarf accreting matter from a low mass companion (usually a late type main sequence star; see review by \citealt{mukai2017}). The accretion flows around the white dwarfs are primarily governed by their magnetic fields. 
For non-magnetic CVs, the accretion disks extends upto the surface of the primary, whereas for polars the magnetic field is strong enough to prevent the formation of the accretion disk. Intermediate Polars (IPs) are a subclass of magnetic CVs, where a partial accretion disk exists and the accretion onto the white dwarf is magnetically funnelled onto their poles from the inner edge of the truncated accretion disks. 
\par
The X-ray emission from these IPs arises from the accreted matter shock heated up to high temperatures (kT $\sim$ 10--50 keV) and must cool before settling onto the surface of the white dwarf. The emergent X-ray spectrum is expected to be the resultant emission from plasmas over a continuous temperature distribution, from the shock temperature to the white dwarf photospheric temperature. High resolution X-ray spectra of CVs are found to be similar to the cooling flows spectra seen in clusters of galaxies \citep{fabian1977,fabian1994}. However, the low temperature emission predicted by the classical
cooling flow models is absent in clusters of galaxies (see, e.g., \citealt{peterson2003}) due to the presence of heating processes, presumably AGN feedback. On the other hand in CVs, cooling continues until the plasma reaches the white dwarf photospheric temperature (T $\sim$ 10$^{4}$ K). The emergent X-rays from this shock heated multi-temperature plasma is modified by the presence of a complex distribution of absorbers in the pre-shock region. This complex absorber is expected to be ionized due to irradiation of the pre-shock flow around the white dwarf. 
For some IPs, a soft component is seen in the X-ray spectra which (in low to medium resolution data) is modelled by a blackbody with a temperature kT $\geq$ 60 eV \citep{evans2007,anzolin2008}. This soft component is thought to arise from reprocessing of the hard X-rays, although there is a considerable uncertainty in the presence of this component.
\par
The X-ray spectra of IPs are often strongly absorbed at a level significantly above the intervening interstellar column. Both the energy dependence of spin modulation amplitudes and the spectral shape of the spin phase-averaged spectra indicate that a simple absorber is not sufficient \citep{norton1989}; these authors used the partial-covering absorber model {\tt pcfabs} available in {\tt XSPEC}\footnote{More recently, the convolution model {\tt partcov} can be used to turn any absorber model into a partial covering absorber model.}. In {\tt pcfabs}, some fraction of the emission is assumed to be observed directly, while the rest (``covering fraction'') is seen through a neutral absorber of column \nh. However, \citet{done1998} pointed out that the partial covering absorber model is far too simplistic for the magnetic CVs. This is because the post-shock region, the source of the X-ray emision, is directly adjacent to the pre-shock flow, the site of the complex absorption. This results in an absorber having a continuous range of \nh, each value with its own differential covering fraction (see Figure 8 of \citealt{mukai2017}). \citet{done1998} proposed to model this using a model in which the differential covering fraction is a power-law function of \nh, and encoded this in an {\tt XSPEC} model {\tt pwab}.
\par
In this paper, we propose a further modification of the {\tt pwab} model, because the accretion flow just above the shock is likely ionized. Photo-ionization model is usually calculated in terms of a cloud of density $n$ at a distance $r$ from the ionizing source with a luminosity of $L$ as $\xi = L / n r^2$. This can be generalized as the ratio of ionizing flux to density, modulo geometrical factor 4$\pi$, when considering an extended cloud neighboring an extended source of ionizing flux. For the pre-shock accretion flow in magnetic CVs, $\xi$ probably exceeds 10 just above the shock but rapidly decreases away from the shock front, due to the combination of geometrical dilution of the ionizing radiation (see Discussion). Observationally, the 0.73 keV edge due to O$_{\rm VII}$ has been detected in V1223~Sgr \citep{mukai2017} and in V2731~Oph \citep{demartino2008}. It is likely that the pre-shock flow is the site for both the ionized absorbers and the photo-ionized emission features originally recognized by \cite{mukai2003} and confirmed here in this work for NY Lup and V1223 Sgr.
\par
The outline of the paper is as follows: in Section 2 we describe the Chandra/HETG and NuSTAR observations of NY Lup and V1223 Sgr. The broad-band X-ray Chandra/HETG and NuSTAR spectra is fitted by the complex absorption model {\tt zxipab} which is described in detail in the Appendix. We discuss these results in Section 3.

\section{Observations and Analysis}

\begin{table}
\centering
\caption{Summary of X-ray observations} 
\label{obs}
\begin{tabular}{c c c c c}
\hline
Source & Telescope & ObsID & Start Time & Exposure (kilosec) \\
\hline
NY Lup & Chandra/ACIS-S HETG & 17874 & 2016-05-23 15:24:24 & 49.41 \\
& Chandra/ACIS-S HETG & 18857 & 2016-05-24 22:04:15 & 32.35 \\
& Chandra/ACIS-S HETG & 18858 & 2016-05-25 22:32:09 & 23.15 \\
& Chandra/ACIS-S HETG & 18859 & 2016-05-28 13:14:07 & 26.71 \\
& NuSTAR & 30001146002 & 2014-08-09 14:51:07 & 23.0 \\
\hline
V 1223 Sgr & Chandra/ACIS-S HETG & 649 & 2000-04-30 16:19:51 & 51.5 \\
           & NuSTAR & 30001144002 & 2014-09-16 02:26:07 & 20.4 \\
\hline
\end{tabular}
\end{table}

V1223 Sgr and NY Lup have been observed by Chandra ACIS-S camera and High Energy Transmission Grating (HETG; \citealt{canizares2005}). HETG consists of two transmission gratings, Medium Energy Grating (MEG) and High Energy Grating (HEG), having absolute wavelength accuracy of 0.0006 \text{\normalfont\AA} and 0.011 \text{\normalfont\AA} respectively. To model the broadband X-ray spectra, we use the NuSTAR observations of V1223 Sgr and NY Lup, which have been previously analysed in \cite{mukai2015} and \cite{hayashi2021}. The Nuclear Spectroscopic Telescope Array (NuSTAR; \citealt{harrison2013}) carries two co-aligned Wolter I telescopes that focus X-rays between 3 and 79 keV onto two independent solid state Focal Plane Modules (FPMA and FPMB). Table 1 gives a summary of the X-ray observations of V1223 Sgr and NY Lup used in this paper.
\par
The Chandra/HETG observations were downloaded from the Chandra archive and were processed by CIAO 4.12 and CALDB 4.9.4. The spectral products, response matrices and ancillary files for different grating orders were obtained by running the CIAO tool {\tt chandra$\_$repro} script. The positive and negative diffraction first order spectra and their corresponding response files were co-added using the CIAO tool {\tt combine$\_$grating$\_$spectra}. The first order spectra extracted from the multiple Chandra/HETG observations of NY Lup was also co-added using CIAO tool {\tt combine$\_$grating$\_$spectra}. 
The NuSTAR data were processed by NuSTARDAS with the latest CALDB files. The source region was defined using a 100'' circular region and the background region was also defined using a 100'' circular region in a source-free region on the same detector. The spectral products, response matrices and ancillary files were obtained by {\tt nuproducts}. The NuSTAR spectra were grouped with {\tt grppha} to have atleast 30 counts per bin. Even though the NuSTAR and Chandra/HETG observations were carried out at different dates, we expect a similarity in the spectral shapes, which is determined by the gravitational potential just above the WD surface \citep{shaw2018}. Therefore, small changes in overall accretion rate lead to small changes in overall normalization without spectral variations, as seen for V2731 Oph \citep{lopes2019}. The changes in the normalization is accounted by the cross-normalization constant of the spectral fits.

\subsection{Simultaneous X-ray spectral fitting of Chandra/HETG and NuSTAR spectra}
\begin{figure*}
\centering
\includegraphics[scale=0.4,angle=-90]{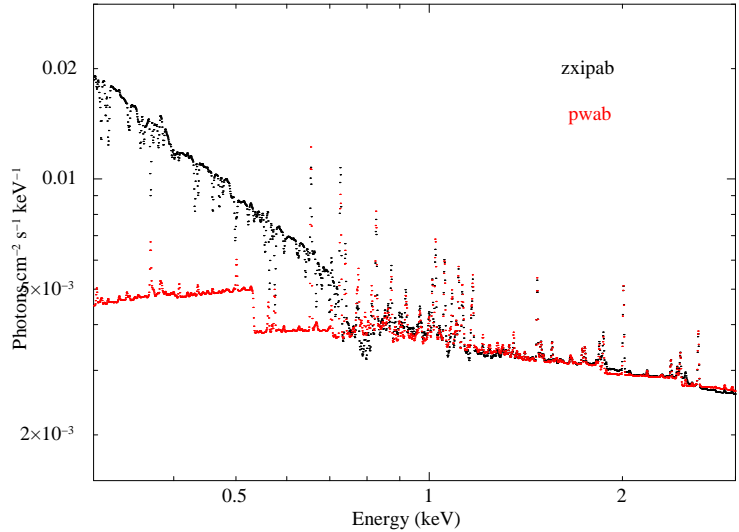}
\caption{Model spectra using {\tt zxipab} and {\tt pwab} as the complex absorption model for V1223 Sgr with a log xi of 0.5.}
\label{model}
\end{figure*}
The post-shock plasma in many mCVs can be well represented by the cooling flow model \citep{done1995, mukai2003}, which contains bremsstrahlung continuum and collisionally excited emission lines from a range of temperatures, with a distribution of emission measures appropriate for isobarically cooling plasma. Along with a multi-temperature plasma and photo-ionized emission lines, there are strong spectral signatures of reflection detected in NY Lup and V1223 Sgr \citep{mukai2015}. \cite{hayashi2021} also studied the reflection in V1223 Sgr using the Suzaku and NuSTAR data and their own model of multi-temperature emission and a reflection component. 
\par
The broadband Chandra/HETG and NuSTAR X-ray spectra of NY Lup and V1223 Sgr are fitted with a cooling flow model for the primary emission ({\tt mkcflow} in {\tt XSPEC}), a reflection model for the spectral signatures of a Compton hump ({\tt reflect} in {\tt XSPEC}) and a complex absorption model. The X-ray spectra, especially at lower energies, is modified by a complex distribution of absorbers in the pre-shock flow, which can be represented as a power-law distribution of neutral absorbers ({\tt pwab} in {\tt XSPEC}) \citep{done1998}. However, the X-ray emission is capable of ionizing the pre-shock flow, hence we expect these complex absorbers to be ionized. Therefore we developed a {\tt zxipab} model as a local model in {\tt XSPEC}, which is a power-law distribution of ionized absorbers. The complex ionized absorber model {\tt zxipab} is similar to {\tt pwab} except the ionization parameter log xi gives an indication of degree of ionization of the absorbers in the pre-shock region. The details of the {\tt zxipab} model are given in the Appendix. We fit the Chandra/HETG spectra of NY Lup and V1223 Sgr with the ionized absorber {\tt zxipab} model and compare it with the fits using a neutral absorber {\tt pwab} model. A gaussian line was used to model the Fe K$\alpha$ line in the spectra, along with a gaussian smoothing model {\tt gsmooth} to account for velocity broadening of emission lines. The power of energy for  sigma variation ($\alpha$) of the {\tt gsmooth} model is fixed to 1.0. In our spectral fits to the data, we do not detect significant interstellar absorption. This may be because the interstellar N$_{H}$ for these objects are too low to be securely detected in the spectral fits, given the complexity of source spectrum and the calibration of the time-dependent low-energy effective area of Chandra/HETG/ACIS-S. The upper limits on interstellar absorption towards NY Lup is 1.4$\times 10^{21}$ cm$^{-2}$ \citep{demartino2008} and V1223 Sgr is 1$\times 10^{21}$ cm$^{-2}$ \citep{beuermann2004}.
\par
In {\tt XSPEC} notation, the model is written as:\\
{\tt constant*complex*gsmooth(reflect*mkcflow+gaussian)} where {\tt complex} is {\tt pwab} or {\tt zxipab}. Figure \ref{model} is the plot of the model spectra using {\tt zxipab} and {\tt pwab} as the complex absorption model for V1223 Sgr with a log xi of 0.5. We see a difference in the prediction between these two complex absorption models at energies below 0.8 keV, along with the presence of absorption lines in the model spectra using an ionized absorber {\tt zxipab} model. Due to limited effective area and energy resolution of Chandra/HETG, we are unable to see these absorption lines in the spectra as well as the difference in the model prediction of fluxes by {\tt zxipab} and {\tt pwab}, especially below 0.5 keV. As shown later in Section 2.2, we see presence of various photo-ionized emission lines in the Chandra/HETG spectra of NY Lup and V1223 Sgr. 

\par
The Chandra/HETG spectra was fitted in energy range 0.5--7.5 keV and NuSTAR/FPMA + FPMB spectra in 3--40 keV. A constant was added for simultaneous Chandra/HETG, NuSTAR/FPMA + FPMB spectral fit to account for cross-normalization difference between the different instruments. We used chi-square statistics for fitting the Chandra/HETG and NuSTAR data in {\tt XSPEC v12.11.0}. The abundance table was used from \cite{asplund2009}. The abundances of the cooling flow model was linked to the reflection model and was allowed to vary during the fit. The lowest temperature to which the plasma cools (kT$_{min}$ of the {\tt mkcflow} model) was fixed to its hard limit of 80.8 eV and the spectrum was calculated by using AtomDB data (switch=2 in {\tt mkcflow} model). The inclination angle of the reflecting surface of the {\tt reflect} model is fixed to cos $\mu$=0.45. The value of spectral parameters obtained by the spectral fits to only NuSTAR data of NY Lup and V1223 Sgr are similar to that reported in \cite{mukai2015}. 
Table 2 gives the spectral parameter of the best fit to the Chandra/HETG and NuSTAR spectral fits using the two complex absorption models: {\tt pwab} and {\tt zxipab}. The errors are calculated at 90\% confidence limit. 
\par
\begin{figure*}
\centering
\includegraphics[scale=0.3]{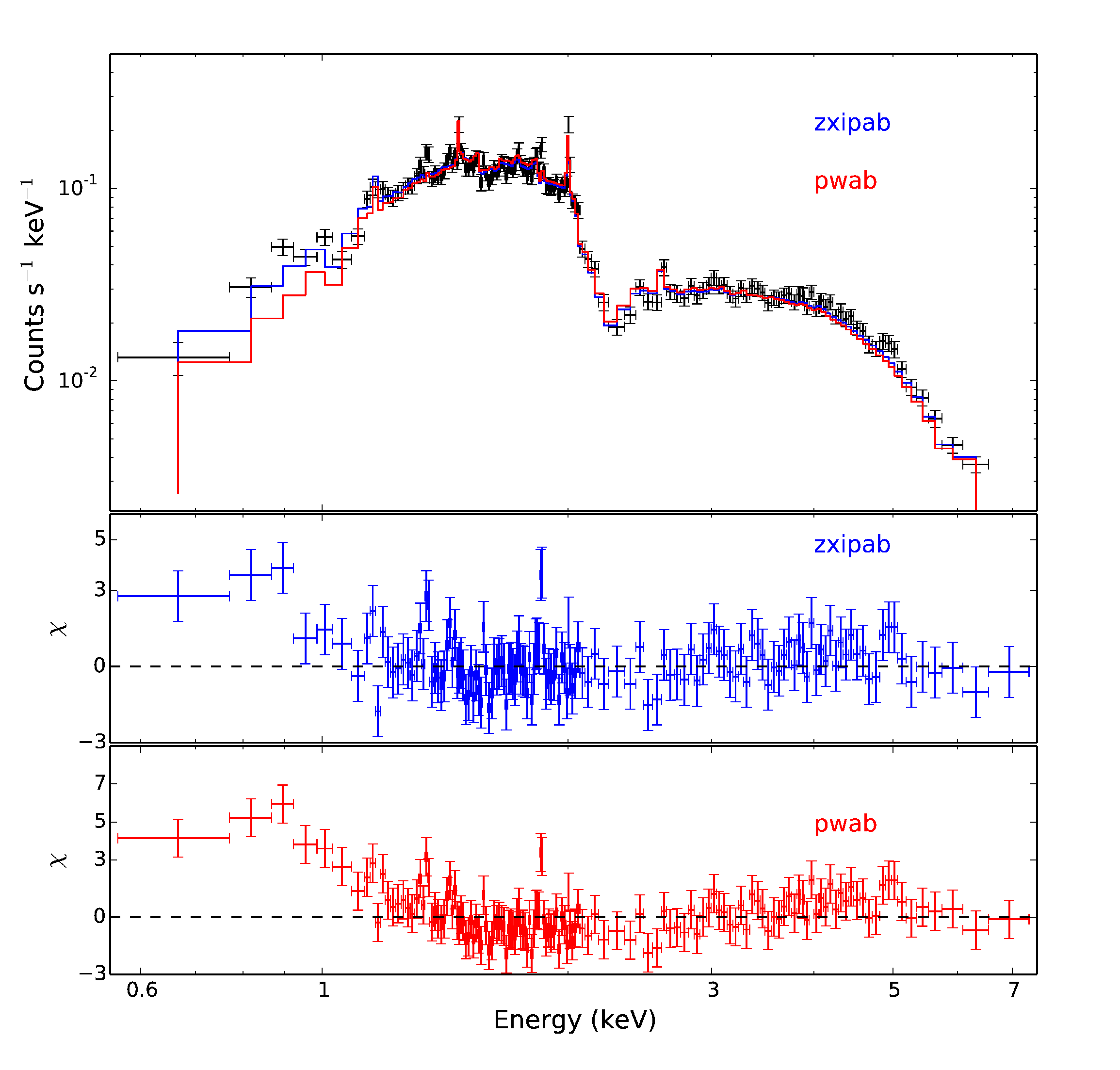}
\includegraphics[scale=0.3]{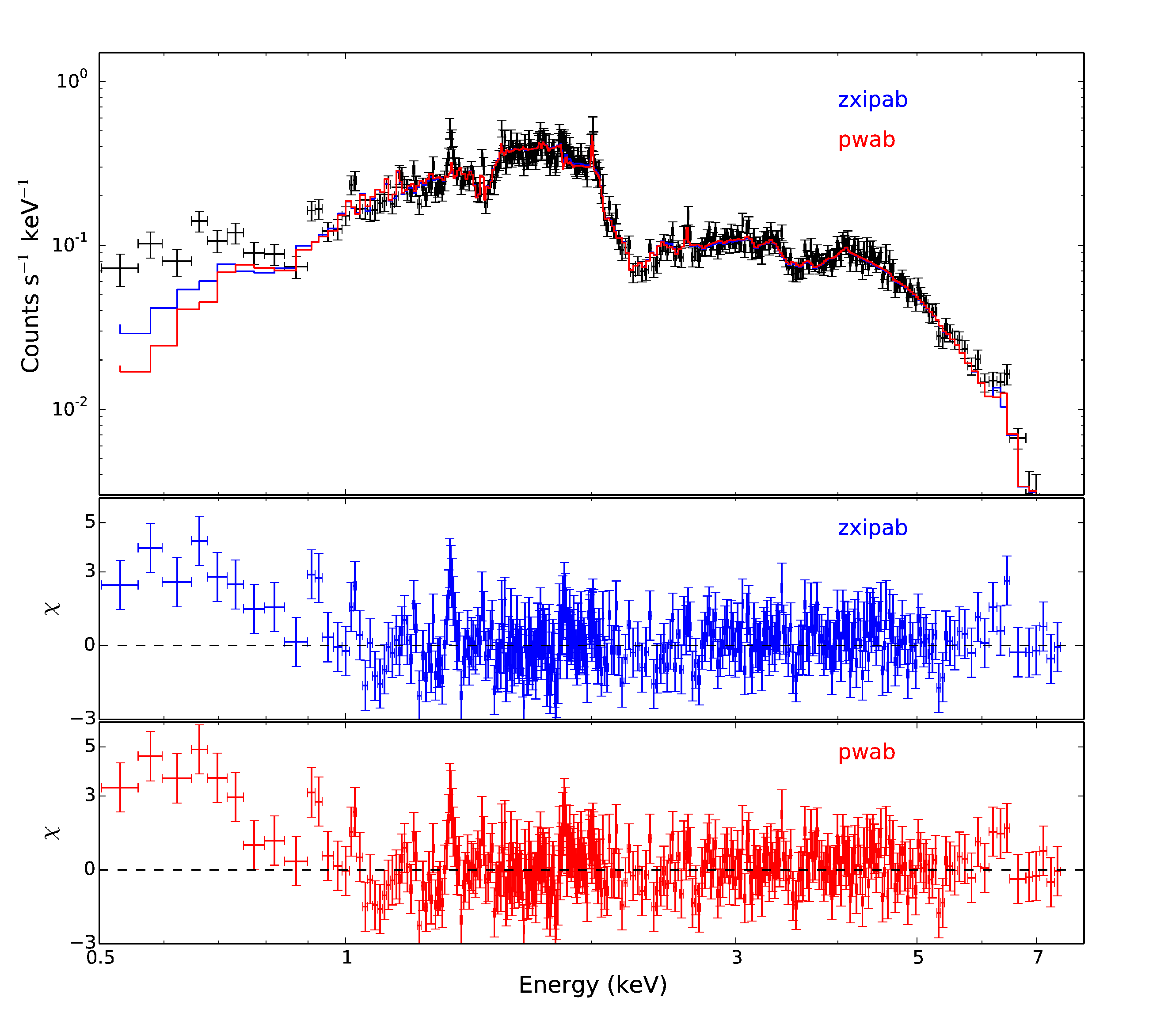}
\caption{Chandra/HETG spectra of NY Lup (left panel) and V1223 Sgr (right panel) overlaid with the complex absorption model {\tt zxipab} (blue) and {\tt pwab} (red).The second and third panels (note the Y scale changes between the two for NY Lup) of the plots are residuals to the {\tt zxipab} and {\tt pwab} models respectively. }
\label{fit_chandra}
\end{figure*}

Figure \ref{fit_chandra} compares the complex absorption models {\tt pwab} and {\tt zxipab} with that of the Chandra/HETG spectra of NY Lup and V1223 Sgr. We see a sharp discontinuity at  the OVII edge around $\sim$ 0.9 keV in the Chandra/HETG spectra, which indicates the presence of warm absorber features \citep{mukai2017}. Although the ionization parameter for NY Lup and V1223 Sgr is not very large, we still expect the pre-shock flow to be ionized due to high X-ray fluxes from these systems. The ionized absorber model {\tt zxipab} is shown to fit the Chandra/HETG data better than the neutral absorber {\tt pwab} model, especially at lower energies. However due to limited sensitivity of Chandra/HETG below 0.5 keV, we do not see a significant change in the fitting statistics between the two complex absorber models as expected from Figure \ref{model}. 
\par
We see residuals to the fit at lower energies in the lower panel of Figure \ref{fit_chandra}. This soft excess is less for the ionized absorber model compared to the neutral absorber model. Some of these residuals are from the photo-ionized emission lines whose fluxes do not match the model predictions of the cooling flow model, described in detail in Section 2.2. We fit this soft excess with a black-body component and find that the estimated temperature is higher than the previously estimated temperature of 60 eV \citep{evans2007,anzolin2008}, which is explained in Discussion. However the F-test shows this black-body is not statistically significant as a fit to this soft excess.

\begin{table*}
\centering
\caption{Best fitting parameter values for the Chandra/HETG + NuSTAR spectra using the model defined in Section 2.1 with a complex absorption model {\tt zxipab} and {\tt pwab}. The errors on the parameters are estimated using 90\% confidence limits}
\begin{tabular}{c c c c c}
\hline
& NY Lup & & V1223 Sgr & \\ 
Parameters & {\tt pwab} & {\tt zxipab} & {\tt pwab} & {\tt zxipab}\\
\hline
C$_{FPMA}$ & 0.94$\pm$0.01 & 0.94$\pm$0.01 & 0.804$\pm$0.005 & 0.805$\pm$0.005 \\
C$_{FPMB}$ & 0.95$\pm$0.01 & 0.94$\pm$0.01 & 0.798$\pm$0.005 & 0.810$\pm$0.006 \\
$\sigma$ (gsmooth; at 6 keV) & 0.011$\pm$0.003 & 0.011$\pm$0.003 & 0.011$\pm$0.004 & 0.011$\pm$0.004 \\
\nh$_{max}$ (10$^{22}$cm$^{-2}$) & 15$^{+2}_{-1}$ & 38$^{+4}_{-3}$ & 15.3$\pm$0.5 & 15.8$\pm$0.7  \\
$\beta$ & -0.80$\pm$0.01 & -0.81$\pm$0.02 & -0.51$\pm$0.01 & -0.46$\pm$0.01 \\
log xi & - & 1.3$\pm$0.2 & - & 0.5$\pm$0.1 \\
rel$\_$ref (reflect) & 1.6$\pm$0.2 & 1.3$^{+0.3}_{-0.1}$ & 0.9$\pm$0.1 & 0.7$\pm$0.1 \\
abundance & 1.1$\pm$0.2  & 1.3$\pm$0.2 & 0.47$\pm$0.05 & 0.45$\pm$0.05 \\
kT$_{max}$ (mkcflow; in keV) & $<$ 73 & $<$ 78 & 41$\pm$2 & 42$\pm$2 \\
Normalization (mkcflow; in M$_\odot$/yr) & (8.9$\pm$0.5) $\times 10^{-10}$ & (9.5$^{+0.8}_{-0.2}$)  $\times 10^{-10}$ & (5.7$\pm$0.2) $\times 10^{-9}$ & (5.9$\pm$0.2) $\times 10^{-9}$ \\
Fe K$\alpha$ (in keV) & 6.43$\pm$0.03  & 6.42$\pm$0.03 & 6.39$\pm$0.02 & 6.39$^{+0.01}_{-0.03}$ \\
Normalization (photons/cm$^{2}$/s) & (1.2$\pm$0.1) $\times 10^{-4}$ & (1.1$\pm$0.1) $\times 10^{-4}$ & (1.2$\pm$0.2) $\times 10^{-4}$ & (1.2$\pm$0.2) $\times 10^{-4}$ \\
$\chi^{2}$ & 1265.46 for 1614 d.o.f & 1261.50 for 1613 d.o.f & 1838.22 for 1898 d.o.f & 1811.80 for 1897 d.o.f \\
Absorbed Flux (0.3-8.0 keV) & 3.1 & 3.2 & 9.8 & 9.7 \\
($\times 10^{-11}$ ergs/sec) & & & & \\
Absorbed Flux (3.0-40.0 keV) & 8.9 & 9.0 & 21.4 & 21.4 \\
($\times 10^{-11}$ ergs/sec) & & & & \\
\hline
\end{tabular}
\end{table*}

\begin{table*}
\centering
\caption{Photo-ionized emission line of NY Lup}
\begin{tabular}{c c c}
\hline
Ion   &  $\lambda$ in \text{\normalfont\AA} & Flux ($\times 10^{-14}$ ergs/sec/cm$^{2}$) \\
\hline
Si XIII i & 6.65 & 1.3$\pm$0.7 \\
Si XIII f & 6.69 & 1.8$\pm$0.7 \\
Mg XI r & 9.18 & 0.9$\pm$0.5 \\
Mg XI i & 9.23 & 1.1$\pm$0.5 \\
Ne IX r & 13.24 & 0.6$\pm$0.2 \\
Ne IX i & 13.45 & 3.6$\pm$0.9 \\
Ne IX f & 13.54 & 4$\pm$1 \\

\hline
\end{tabular}
\end{table*}

\begin{table*}
\centering
\caption{Photo-ionized emission line of V1223 Sgr}
\begin{tabular}{c c c}
\hline
Ion   &  $\lambda$ in \text{\normalfont\AA} & Flux ($\times 10^{-14}$ ergs/sec/cm$^{2}$) \\
\hline
Mg XI r & 9.18 & 2$\pm$1 \\
Mg XI i & 9.23 & 4$\pm$1 \\
Mg XI f & 9.34 & 1.1$\pm$0.9 \\
Ne IX i & 13.45 & 3$\pm$2 \\
Ne IX f & 13.54 & 1.4$\pm$0.9 \\
OVII i & 21.61 & 2.1$\pm$0.9 \\
OVII f & 21.79 & 1.3$\pm$0.9 \\

\hline
\end{tabular}
\end{table*}

\subsection{Photo-ionized emission lines}
\begin{figure*}
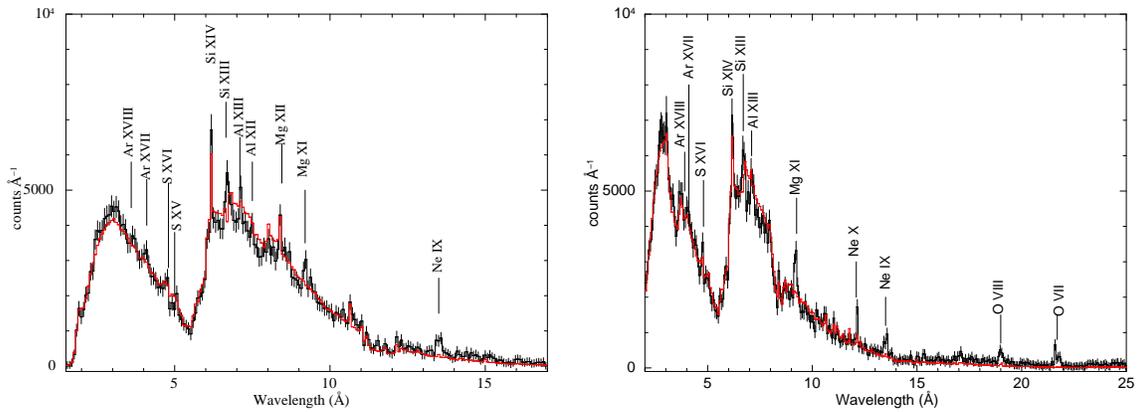

\centering
\includegraphics[scale=0.3,angle=-90]{nylup_meg_line.ps}
\includegraphics[scale=0.3,angle=-90]{v1223sgr_meg_line.ps}
\caption{Chandra MEG summed first order spectra of NY Lup (left panel) and V1223 Sgr (right panel), along with various emission lines. The red line is the best-fit spectral model using an ionized absorber model {\tt zxipab} and cooling flow model {\tt mkcflow}, along with a reflection component {\tt reflect}. Data show excess flux for Si XIII, Mg XI and Ne IX lines for NY Lup and Mg XI, Ne IX and OVII lines for V1223 Sgr}.
\label{lines}
\end{figure*}

\begin{figure*}
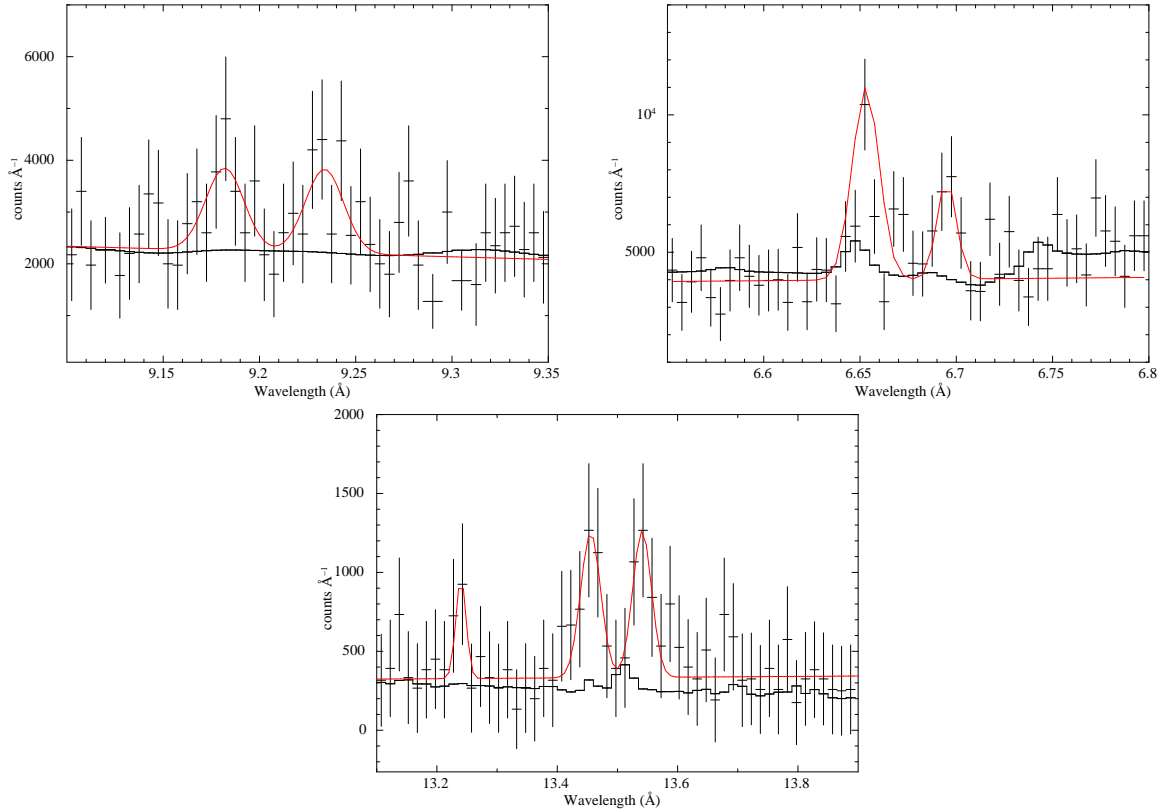

\centering
\includegraphics[scale=0.3,angle=-90]{NYLup_MgXI.ps}
\includegraphics[scale=0.3,angle=-90]{NYLup_Si_XIII.ps}
\includegraphics[scale=0.3,angle=-90]{NYLup_Ne_IX.ps}
\caption{Chandra MEG summed first order spectra of NY Lup showing the presence of He like triplets of Mg XI (left panel), Si XIII (right panel) and Ne IX (bottom panel). The gaussian fits to the emission lines are shown in red and the predictions of the line strengths with {\tt mkcflow} model is shown in black line. The spectra is binned for clarity}
\label{nylup_line}
\end{figure*}

\begin{figure*}
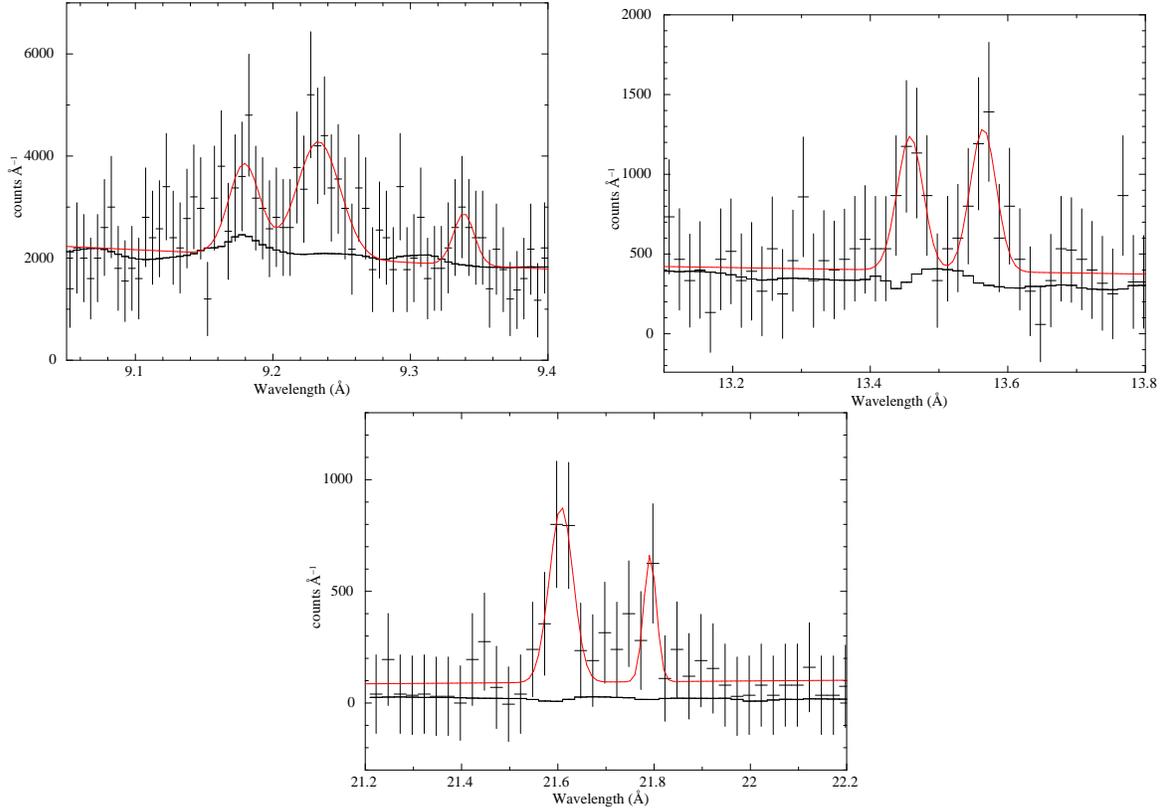

\centering

\includegraphics[scale=0.3,angle=-90]{V1223Sgr_Mg_XI.ps}
\includegraphics[scale=0.3,angle=-90]{V1223Sgr_Ne_IX.ps}
\includegraphics[scale=0.3,angle=-90]{V1223Sgr_O_VII.ps}
\caption{Chandra MEG summed first order spectra of V1223 Sgr showing the presence of He like triplets of Mg XI (left panel), Ne IX (middle panel) and O VII (right panel). The gaussian fits to the emission lines are shown in red and the predictions of the line strengths with {\tt mkcflow} model is shown in black line. The spectra is binned for clarity}
\label{v1223sgr_line}
\end{figure*}

\begin{figure*}
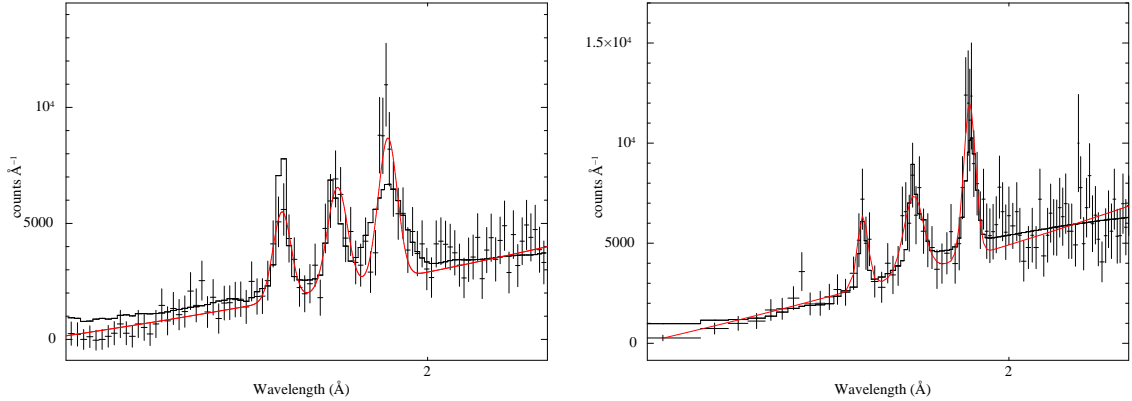

\centering
\includegraphics[scale=0.3,angle=-90]{NYLup_Fe.ps}
\includegraphics[scale=0.3,angle=-90]{V1223Sgr_Fe.ps}
\caption{Chandra HEG summed first order spectra of NY Lup (left panel) and V1223 Sgr (right panel) showing the presence of neutral, He like (Fe XXV) and H like Fe K$\alpha$ lines (Fe XXVI). The gaussian fits to the emission lines are shown in red and the predictions of the line strengths with {\tt mkcflow} model plus a gaussian line to model the Fe K$\alpha$ line, is shown in black line. The spectra is binned for clarity}
\label{Fe_line}
\end{figure*}

Figure \ref{lines} shows the MEG summed first order spectra of NY Lup and V1223 Sgr, displaying the emission lines along with their predictions from the cooling flow {\tt mkcflow} model. We see the presence of H and He like ions emission lines from Ne, Mg, Al, Si, Ar and S in the MEG spectra of NY Lup and V 1223 Sgr. We also see emission lines corresponding to He and He like ion of O in V1223 Sgr, which are not seen in NY Lup, most likely due to the decreased low energy effective area of ACIS in the 2016 Chandra HETG observation of NY Lup. We find the model predictions of the line strengths from the {\tt mkcflow} model describe the fluxes of most of the emission lines, except for certain He like triplets of Si, Mg XI, Ne IX and O VII. Figure \ref{nylup_line} shows the MEG first order summed spectra of NY Lup around Mg XI, Si XIII and Ne IX emission lines. We can resolve He like triplet lines of Ne IX emission lines, which are fitted with three gaussian lines in the plot. For Mg XI and Si XIII, we can only resolve two lines out of the expected He like triplets, due to low statistics of the third line. These are fitted with two gaussian lines in the plot. Figure \ref{v1223sgr_line} shows the MEG first order summed spectra of V1223 Sgr around Mg XI, Ne IX and O VII emission lines. We can resolve the He like triplet lines of Mg XI and are fitted with three gaussian lines in the plot. For Ne IX and O VII, we can only resolve two lines out of the expected He like triplets, due to low statistics of the third line. These are fitted with two gaussian lines in the plot. In all these plots, the primary emission is fitted by a first order polynomial plus the gaussians for the emission lines. To estimate the excess X-ray fluxes of these emission lines which are resolved into double or triple lines, we fit them with a gaussian line and estimate their excess flux over the predictions from the {\tt mkcflow} model. Table 3 and 4 shows the excess X-ray flux of these photo-ionized lines over their flux predictions by the {\tt mkcflow} model. The errors are calculated at 90\% of the confidence limit on the normalisation of the gaussian lines used to model these emission lines and propagated to account for the errors on the fluxes. 
\par 
Figure \ref{Fe_line} shows the HEG summed first order spectra around Fe lines of NY Lup and V1223 Sgr, with the emission lines corresponding to neutral Fe (Fe K$\alpha$), He like Fe (Fe XXV) and H like Fe (Fe XXVI). The primary emission is modelled with the cooling flow model {\tt mkcflow} and a gaussian line for Fe K$\alpha$. 

\begin{figure*}
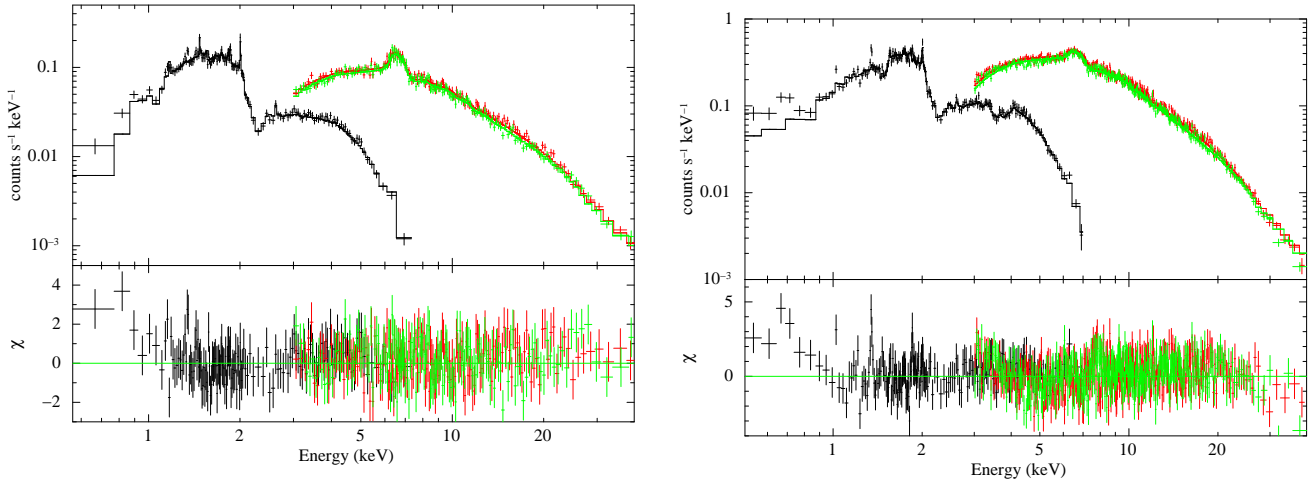

\centering
\includegraphics[scale=0.35,angle=-90]{NYLup_zxipab_emission.ps}
\includegraphics[scale=0.35,angle=-90]{V1223Sgr_zxipab_emission.ps}
\caption{Chandra/HETG (black) + NuSTAR (FPMA: red and FPMB: green) spectral fit to NY Lup (left panel) and V1223 Sgr (right panel) using the model defined in Section 2.1 with the ionized absorber model {\tt zxipab} (right panel), along with ratio of the data to the model. The Chandra/HETG and NuSTAR spectra are rebinned for clarity.}
\label{fit_zxipab}
\end{figure*}

Our complete fitting results using a spectral model with an ionized absorber model {\tt zxipab}, cooling flow model {\tt mkcflow}, reflection component {\tt reflect} and gaussian emission lines for Fe K$\alpha$ line and the emission lines described in Section 2.2 whose observed fluxes do not match the cooling flow model predictions, are shown in Figure \ref{fit_zxipab} along with their residuals for NY Lup and V1223 Sgr.

\section{Discussion}

Previous study by \cite{mukai2003} classified the Chandra/HETG spectra of 7 CVs into two groups based on their X-ray spectra. EX Hya, V603 Aql, U Gem and SS Cyg were classified as cooling-flow type since their X-ray spectra are described well using the {\tt mkcflow} model, which was originally developed to model the cooling flows in clusters of galaxies \citep{mushotzky1988}. In these four CVs, the cooling flow model provides a good explanation for the continuum shape as well as the strong H and He like emission lines of O, Ne, Mg, Al, Si, S and entire Fe L complex (Figure 1 in \citealt{mukai2003}). On the other hand, in the three other CVs studied by \cite{mukai2003} (V1223 Sgr, AO Psc, and GK Per), the continuum is much harder than what the cooling flow model {\tt mkcflow} predicts; the details of the observed emission lines are also different from {\tt mkcflow} predictions. Specifically, the Fe L lines are weaker than expected, and the ratio of fluxes of of He-like ions to those of H-like ions are higher than predicted. These discrepancy led \cite{mukai2003} to explain the entire Chandra/HETG spectra of these three CVs using a photo-ionization model.
\par
However, there are several issues with this interpretation. First, it is difficult to imagine a situation where the modest luminosity (of order 10$^{34}$ erg/s) of these CVs can nevertheless produce H-like and He-like lines of Fe through photo-ionization. Second, the hard power-law continuum of \cite{mukai2003} was an ad-hoc assumption, incompatible with the standard understanding that the strong shock is the primary source of emission in IPs. Various studies (including \citealt{mukai2015}) focusing on the harder end of the X-ray spectra of these `photo-ionized' CVs have already demonstrated that they can be modelled by a cooling flow spectra modified by a complex distribution of absorbers in the pre-shock region. Such models attribute the H-like and He-like emission lines of Fe to the collisionally-excited, multi-temperature, post-shock plasma. They explain the hard power-law like continuum as that of multi-temperature bremsstrahlung emission hardened by the complex absorber (see Appendix for an analytical understanding of this). 
\par
Given the specific accretion rate (accretion rate per unit surface area) $\dot m$, and assuming a shock very close to the stellar surface, we can estimate both the density of the pre-shock flow and the photon flux through this region. The pre-shock density is $\dot m$ divided by the free-fall velocity. For a 0.8 M$_\odot$ white dwarf, the free-fall velocity is 5.5$\times 10^8$ cm\,s$^{-1}$ and the pre-shock density is 1.82$\times 10^{-9}$ $\dot m$ g\,cm$^{-3}$, or number density of 1.8$\times 10^{15}$ $\dot m$ cm$^{-3}$.
We can obtain the accretion luminosity per unit area by multiplying $\dot m$ with gravitational potential (GM/R) of the white dwarf. For a 0.8 M$_\odot$ white dwarf, the luminosity per unit area is 1.5$\times 10^{17}$ $\dot m$ erg\,s$^{-1}$, of which half will be radiated downward and converted into soft photons, so the hard X-ray luminosity is 7.5$\times 10^{16}$ $\dot m$ erg\,s$^{-1}$. The ionization parameter, $\xi = L / n r^2$, generalized for a cloud adjacent to an extended source of ionizing radiation, is independent of $\dot m$ and $\sim$40 in the 0.8 M$_\odot$ immediately above the shock. However, the actual value will be lower due to geometrical dilution (both due to finite shock height, and finite distance above the shock front). We therefore expect the immediate pre-shock flow to be modestly, but measurably ionized. This expectation is confirmed by the presence of photo-ionized emission lines and warm absorption features, seen in Figure \ref{fit_chandra} and \ref{lines}. This justifies our decision to develop a model for ionized distribution of absorbers in the pre-shock region, {\tt zxipab} (see Appendix for the details of the model).
\par
Figure \ref{model} shows the model prediction of the complex absorption models {\tt zxipab} and {\tt pwab} at lower energies. There is a significant deviation between the prediction of the two models at energies $\leq$ 0.8 keV. The ionized absorber model {\tt zxipab} shows the presence of sharp discontinuity due to OVII edge at $\sim$ 0.9 keV, which is also seen in the Chandra/HETG spectrum of V1223 Sgr (Figure \ref{fit_chandra}). Due to the decreased effective area of Chandra/HETG, this is not as obvious for NY Lup. When we see the X-rays escaping through the pre-shock region, it shows warm absorber features. This pre-shock region is likely the origin of the photo-ionized emission lines as well. Future missions like XRISM would be important in detecting the presence of warm absorber like absorption lines from these CVs. 
\par
The Chandra/HETG spectra of NY Lup and V1223 Sgr in Figure \ref{lines} shows the presence of emission lines from H and He like ions of Ne, Mg, Al, Si, S, Fe. Many of these lines are from the collisionally excited plasma, since the cooling flow model {\tt mkcflow} predict the line fluxes accurately. However the line predictions of the cooling flow model fails to account for the He like lines of some medium Z elements. These are the lines that we ascribe to photo-ionization. He-like triplets have a good diagnostic capability, and line ratios can be used as an independent test of the line origin (collisional vs photo-ionized) \citep{porquet2000, porquet2001}. 
However, \cite{mukai2003} argued that this diagnostic is not appropriate for CVs, because the radial velocity gradient of thousands of km/s can result in all the lines remaining completely unsaturated at all ionic column densities. This leaves the plasma in photo-excitation dominated regime, in which the resonance lines are stronger than in photo-ionization dominated regime \citep{kinkhabwala2002}. Also, resonant scattering may have an important effect on the line ratios in CVs \citep{terada2004}. Moreover, Chandra HETG observations of CVs are far too underexposed to be able to use such diagnostics \citep{schlegel2014}. For all these reasons, existing data cannot be used to derive a definitive origin of the lines that the cooling flow models cannot account for. The neutral Fe K$\alpha$ originates both from the white dwarf surface and pre-shock flow \citep{ezuka1999}. 
\par
The NuSTAR observations of NY Lup and V1223 Sgr show the presence of a Compton hump in its hard X-ray spectra and an energy dependent spin modulations with an amplitude of $\leq 10 \%$ \citep{mukai2015}. Both these findings point to the fact that the shock is close to the white dwarf surface (a shock height of 0.05 R$_{wd}$ for V1223 Sgr and a negligible shock height for NY Lup; see \citealt{mukai2015}) resulting in strong Compton reflection off the white dwarf surface. This also results in strong complex absorption, and significant degree of ionization of the pre-shock flow. All these are consequences of high specific accretion rate $\dot m$. In contrast, EX Hya has a lower $\dot m$ and hence tall shocks ($\sim$ R$_{wd}$; where $R_{wd}$ is radius of the white dwarf ) allowing X-rays to escape the post-shock region by travelling through the sides. Hence there is an absence of complex absorption or reflection in its X-ray spectra \citep{luna2018}. 
\par
Figure \ref{fit_zxipab} shows the broadband fit of the Chandra/HETG and NuSTAR spectra of NY Lup and V1223 Sgr in energy range of 0.5--40 keV. The spectral model used for this broadband fit consists of the ionized absorber model {\tt zxipab}, cooling flow model {\tt mkcflow}, reflection component {\tt reflect} and gaussian lines to model the emission from Fe K$\alpha$ and photo-ionized lines. The presence of every model component is physically motivated. We therefore conclude that this is superior to the artificial model of \cite{mukai2003}.
\par
Given that the shock is close to the white dwarf surface in many magnetic CVs, roughly half the radiation will intercept the stellar surface and heat it up. For a 0.8 M$_\odot$ white dwarf accreting at $\dot m$ of 1.0 g\,s$^{-1}$cm$^{-2}$, kT$\sim$20 eV blackbody is expected. While just such a component is prominent in the majority of polars, no such component was seen in IPs until \cite{haberl1995} discovered three ``soft IPs'' with such a component. \cite{haberl2002} inferred the presence of such a component in NY~Lup, with an estimated temperature of 86 eV. Since then, the presence of the soft component has been inferred in quite a few IPs \citep{evans2007,anzolin2008}. However, the high inferred blackbody temperatures in several IPs, including NY~Lup, presents a potentially serious problem, because the luminosity per unit area is so high that some may violate the Eddington limit, and is likely to violate the atmospheric limit \citep{williams1987}. However previous claims of a soft component is based on the spectral fits without ionized complex absorbers or photo-ionized emission lines.
In Figure \ref{fit_chandra}, we find residuals at lower energies while fitting the Chandra/HETG spectra with complex absorber models; the residuals are lower for the {\tt zxipab} model compared to the {\tt pwab} model. However there are uncertainties related to the interstellar absorption related to its estimation using Chandra/HETG data and the instrumental calibrations issues related to ACIS degradation, which could be related to these soft excess.  We attempted to fit these soft excess with a blackbody component. However, this did not result in a statistically significant improvement of fit quality, suggesting the soft excess may not have the right shape to be explained with a blackbody component. The temperature we derive, which is unphysical for these systems, is therefore also a suspect. We tentatively conclude that, when the correct spectral model (including ionized complex absorber and photo-ionized lines), the need for the blackbody may disappear, or the blackbody temperature may not be as hot as previously claimed.

\section{Summary}

We have developed a new, ionized version ({\tt zxipab}) of the complex absorber model {\tt pwab}, and applied it, along with the cooling flow model {\tt mkcflow}, the reflection model {\tt reflect}, and Gaussian emission lines, to Chandra/HETG and NuSTAR spectra of two IPs, NY Lup and V1223 Sgr. Our findings are as follows:
\begin{itemize}
    \item The ionized version provides a better, although still not a complete explanation of the spectra below 1 keV than the neutral version.
    \item We attribute the excess emission lines that cannot be explained with the absorbed cooling flow as due to photo-ionized emission from the immediate pre-shock flow, the same region that is responsible for the complex absorption.
    \item Once both these are taken into account, the soft blackbody component that was previously claimed for NY Lup may not be necessary. The previously claimed parameters of the blackbody component certainly are subject to revision.
\end{itemize}

\section*{Appendix: Complex Ionized Absorber Model, zxipab}
\subsection*{Analytical Considerations}
We take equation (1) of \cite{done1998} as our starting point.

\begin{equation}
S(E) = S_{\rm int}(E) A \int_{N_{H,min}}^{N_{H,max}} N_H^\beta \exp{(-N_H \sigma
(E))} dN_H
\end{equation}

This assumes a power-law distribution of covering fraction as a function
of the absorbing column, i.e., $C_f(N_H) \propto N_H^\beta$ and
$A$ is the normalization such that
$\int_{N_{H,min}}^{N_{H,max}} C_f(N_H)=1.0$. The effect of such an absorber
is markedly different from that of a simple absorber. The latter results
in an exponential decline in transmission at low energies, which is often
the handle that allows one to fit for \nh. The combination of a simple
absorber (e.g., from the inter-stellar medium) and a partial covering absorber produces two such exponential cut-offs, with a region in energy where the observed spectrum is dominated by the unabsorbed component. The
effect of of {\tt pwab} is more power-law like (see, e.g., Figure 9 of
\citealt{mukai2017}).

This can be understood using the following considerations. By substituting
$u=N_H \sigma(E)$, we obtain

\begin{equation}
T(E) \propto \sigma(E)^{-\beta-1} \int_{u_{min}}^{u_{max}} u^\beta \exp{(-u)} du
\end{equation}

for the transmission $T(E)$.  The integral can be expressed in terms of
incomplete gamma function
$P(a,x)=1/\Gamma(a) \int_0^x t^{a-1} \exp{(-t)} dt$ so that

\begin{equation}
T(E) \propto \sigma(E)^{-\beta-1} [P(\beta+1, u_{max}) - P(\beta+1, u_{min})]
\end{equation}

The properties of the incomplete gamma function is well known: for a small
positive value of $a$, it rises quickly from 0 to $\sim$1.  If $N_{H,min}$
is sufficiently small and $N_{H,max}$ is sufficiently large, there is
a range of $E$ such that $P(\beta+1, u_{max}) \sim 1$ and
$P(\beta+1, u_{min}) \sim 0$, so that

\begin{equation}
T(E) \propto  \sigma(E)^{-\beta-1}
\end{equation}

At the same time, $\sigma(E)$ itself can be approximated by a power law
of $E$.  For example, 

\begin{equation}
\sigma(E) = 2.242 \times 10^{-22} (E/1 keV)^{-8/3} cm^{-2}
\end{equation}

is a good approximation of the photoelectric cross section of
\citet{morrison1983} between O and Fe edges.  Thus,
the power-law distribution of covering fraction can lead to
a power-law like, rather than exponential, low-energy cut-off.

$\sigma$ has discontinuities at the major edges. The neutral
and ionized versions are different in the positions and the
depth of the discontinuities. In particular, if all K shell
electrons of lower Z elements are ionized, $\sigma$ becomes
close to 0 at low energies. In such a situation, the use of
{\tt zxipab} is necessary, while a combination of {\tt pwab} and
an edge would provide a poor approximation for energies where
$\sigma \sim 0$.

\subsection*{The implementation and current limitations}

The neutral version of the model, {\tt pwab}, numerically integrates
the neutral absorber model, {\tt wabs} \citep{morrison1983}.
For {\tt zxipab}, we replace
the call to {\tt wabs} model with a pre-calculated grid of XSTAR
photoionization model that was originally used for the partial covering
warm absorber model, {\tt zxipcf} \citep{reeves2008}. Model
parameters are the minimum and the maximum \nh, power law index for
covering fraction (from {\tt pwab}), the ionization parameter log($\xi$),
and the redshift of the absorber.

As with {\tt pwab}, {\tt zxipab} approximates the true differential
covering fraction distribution using a power law. This is an intrinsic
limitation of the model. A more realistic model of pre-shock flow
absorption in magnetic CVs should replace the power law with a more
realistic function, although this could rapidly expand the number of
free parameters to describe, e.g., the shape of the footpoint of the
accretion curtain. Also, {\tt zxipab} assumes a single log($\xi$)
for the absorber, whereas a realistic absorber model would also include
a distribution of log($\xi$) as a function of distance from the shock
front. Finally, {\tt zxipab} assumes the absorber with solar abundances,
which is not guaranteed to hold for all magnetic CVs, particularly
for the nitrogen abundance (see, e.g., the case of V2731~Oph:
\citealt{lopes2019}).
\par
{\tt zxipab} is available as a new model in XSPEC through this website\footnote{https://heasarc.gsfc.nasa.gov/docs/xanadu/xspec/newmodels.html}.

\section*{Acknowledgement}
We thank the anonymous referee for helpful comments. The scientific results reported here are based on observations made by the Chandra X-ray observatory. Support for this work was provided by the NASA through Chandra Grant Number GO6-17025A issued by the Chandra X-ray Observatory Center, which is operated by the Smithsonian Astrophysical Observatory for and on behalf of the NASA under contract NAS8-03060. This research has made use of software provided by the Chandra X-ray Center (CXC). We thank Chris Done for her scientific insight and technical assistance regarding {\tt pwab} and {\tt zxipcf} that allowed us to combine the two into {\tt zxipab}. We also thank Keith Arnaud for his assistance in the {\tt XSPEC} implementation of {\tt zxipab}.

\software{CIAO (v4.10; \citealt{fruscione2006}), XSPEC (v12.11.0; \citealt{arnaud1996}).}


\begin{thebibliography}{}
\expandafter\ifx\csname natexlab\endcsname\relax\def\natexlab#1{#1}\fi

\bibitem[{{Anzolin} {et~al.}(2008){Anzolin}, {de Martino}, {Bonnet-Bidaud},
  {Mouchet}, {G{\"a}nsicke}, {Matt}, \& {Mukai}}]{anzolin2008}
{Anzolin}, G., {de Martino}, D., {Bonnet-Bidaud}, J.~M., {et~al.} 2008, \aap,
  489, 1243

\bibitem[{{Arnaud}(1996)}]{arnaud1996}
{Arnaud}, K.~A. 1996, in Astronomical Society of the Pacific Conference Series,
  Vol. 101, Astronomical Data Analysis Software and Systems V, ed. G.~H.
  {Jacoby} \& J.~{Barnes}, 17

\bibitem[{{Asplund} {et~al.}(2009){Asplund}, {Grevesse}, {Sauval}, \&
  {Scott}}]{asplund2009}
{Asplund}, M., {Grevesse}, N., {Sauval}, A.~J., \& {Scott}, P. 2009, \araa, 47,
  481

\bibitem[{{Beuermann} {et~al.}(2004){Beuermann}, {Harrison}, {McArthur},
  {Benedict}, \& {G{\"a}nsicke}}]{beuermann2004}
{Beuermann}, K., {Harrison}, T.~E., {McArthur}, B.~E., {Benedict}, G.~F., \&
  {G{\"a}nsicke}, B.~T. 2004, \aap, 419, 291

\bibitem[{{Canizares} {et~al.}(2005){Canizares}, {Davis}, {Dewey}, {Flanagan},
  {Galton}, {Huenemoerder}, {Ishibashi}, {Markert}, {Marshall}, {McGuirk},
  {Schattenburg}, {Schulz}, {Smith}, \& {Wise}}]{canizares2005}
{Canizares}, C.~R., {Davis}, J.~E., {Dewey}, D., {et~al.} 2005, \pasp, 117,
  1144

\bibitem[{{de Martino} {et~al.}(2008){de Martino}, {Matt}, {Mukai},
  {Bonnet-Bidaud}, {Falanga}, {G{\"a}nsicke}, {Haberl}, {Marsh}, {Mouchet},
  {Littlefair}, \& {Dhillon}}]{demartino2008}
{de Martino}, D., {Matt}, G., {Mukai}, K., {et~al.} 2008, \aap, 481, 149

\bibitem[{{Done} \& {Magdziarz}(1998)}]{done1998}
{Done}, C., \& {Magdziarz}, P. 1998, \mnras, 298, 737

\bibitem[{{Done} {et~al.}(1995){Done}, {Osborne}, \& {Beardmore}}]{done1995}
{Done}, C., {Osborne}, J.~P., \& {Beardmore}, A.~P. 1995, \mnras, 276, 483

\bibitem[{{Evans} \& {Hellier}(2007)}]{evans2007}
{Evans}, P.~A., \& {Hellier}, C. 2007, \apj, 663, 1277

\bibitem[{{Ezuka} \& {Ishida}(1999)}]{ezuka1999}
{Ezuka}, H., \& {Ishida}, M. 1999, \apjs, 120, 277

\bibitem[{{Fabian}(1994)}]{fabian1994}
{Fabian}, A.~C. 1994, \araa, 32, 277

\bibitem[{{Fabian} \& {Nulsen}(1977)}]{fabian1977}
{Fabian}, A.~C., \& {Nulsen}, P.~E.~J. 1977, \mnras, 180, 479

\bibitem[{{Fruscione} {et~al.}(2006){Fruscione}, {McDowell}, {Allen},
  {Brickhouse}, {Burke}, {Davis}, {Durham}, {Elvis}, {Galle}, {Harris},
  {Huenemoerder}, {Houck}, {Ishibashi}, {Karovska}, {Nicastro}, {Noble},
  {Nowak}, {Primini}, {Siemiginowska}, {Smith}, \& {Wise}}]{fruscione2006}
{Fruscione}, A., {McDowell}, J.~C., {Allen}, G.~E., {et~al.} 2006, in Society
  of Photo-Optical Instrumentation Engineers (SPIE) Conference Series, Vol.
  6270, Society of Photo-Optical Instrumentation Engineers (SPIE) Conference
  Series, ed. D.~R. {Silva} \& R.~E. {Doxsey}, 62701V

\bibitem[{{Haberl} \& {Motch}(1995)}]{haberl1995}
{Haberl}, F., \& {Motch}, C. 1995, \aap, 297, L37

\bibitem[{{Haberl} {et~al.}(2002){Haberl}, {Motch}, \& {Zickgraf}}]{haberl2002}
{Haberl}, F., {Motch}, C., \& {Zickgraf}, F.~J. 2002, \aap, 387, 201

\bibitem[{{Harrison} {et~al.}(2013){Harrison}, {Craig}, {Christensen},
  {Hailey}, {Zhang}, {Boggs}, {Stern}, {Cook}, {Forster}, {Giommi},
  {Grefenstette}, {Kim}, {Kitaguchi}, {Koglin}, {Madsen}, {Mao}, {Miyasaka},
  {Mori}, {Perri}, {Pivovaroff}, {Puccetti}, {Rana}, {Westergaard}, {Willis},
  {Zoglauer}, {An}, {Bachetti}, {Barri{\`e}re}, {Bellm}, {Bhalerao},
  {Brejnholt}, {Fuerst}, {Liebe}, {Markwardt}, {Nynka}, {Vogel}, {Walton},
  {Wik}, {Alexander}, {Cominsky}, {Hornschemeier}, {Hornstrup}, {Kaspi},
  {Madejski}, {Matt}, {Molendi}, {Smith}, {Tomsick}, {Ajello}, {Ballantyne},
  {Balokovi{\'c}}, {Barret}, {Bauer}, {Blandford}, {Brandt}, {Brenneman},
  {Chiang}, {Chakrabarty}, {Chenevez}, {Comastri}, {Dufour}, {Elvis}, {Fabian},
  {Farrah}, {Fryer}, {Gotthelf}, {Grindlay}, {Helfand}, {Krivonos}, {Meier},
  {Miller}, {Natalucci}, {Ogle}, {Ofek}, {Ptak}, {Reynolds}, {Rigby},
  {Tagliaferri}, {Thorsett}, {Treister}, \& {Urry}}]{harrison2013}
{Harrison}, F.~A., {Craig}, W.~W., {Christensen}, F.~E., {et~al.} 2013, \apj,
  770, 103

\bibitem[{{Hayashi} {et~al.}(2021){Hayashi}, {Kitaguchi}, \&
  {Ishida}}]{hayashi2021}
{Hayashi}, T., {Kitaguchi}, T., \& {Ishida}, M. 2021, \mnras, 504, 3651

\bibitem[{{Kinkhabwala} {et~al.}(2002){Kinkhabwala}, {Sako}, {Behar}, {Kahn},
  {Paerels}, {Brinkman}, {Kaastra}, {Gu}, \& {Liedahl}}]{kinkhabwala2002}
{Kinkhabwala}, A., {Sako}, M., {Behar}, E., {et~al.} 2002, \apj, 575, 732

\bibitem[{{Lopes de Oliveira} \& {Mukai}(2019)}]{lopes2019}
{Lopes de Oliveira}, R., \& {Mukai}, K. 2019, \apj, 880, 128

\bibitem[{{Luna} {et~al.}(2018){Luna}, {Mukai}, {Orio}, \& {Zemko}}]{luna2018}
{Luna}, G.~J.~M., {Mukai}, K., {Orio}, M., \& {Zemko}, P. 2018, \apjl, 852, L8

\bibitem[{{Morrison} \& {McCammon}(1983)}]{morrison1983}
{Morrison}, R., \& {McCammon}, D. 1983, \apj, 270, 119

\bibitem[{{Mukai}(2017)}]{mukai2017}
{Mukai}, K. 2017, \pasp, 129, 062001

\bibitem[{{Mukai} {et~al.}(2003){Mukai}, {Kinkhabwala}, {Peterson}, {Kahn}, \&
  {Paerels}}]{mukai2003}
{Mukai}, K., {Kinkhabwala}, A., {Peterson}, J.~R., {Kahn}, S.~M., \& {Paerels},
  F. 2003, \apjl, 586, L77

\bibitem[{{Mukai} {et~al.}(2015){Mukai}, {Rana}, {Bernardini}, \& {de
  Martino}}]{mukai2015}
{Mukai}, K., {Rana}, V., {Bernardini}, F., \& {de Martino}, D. 2015, \apjl,
  807, L30

\bibitem[{{Mushotzky} \& {Szymkowiak}(1988)}]{mushotzky1988}
{Mushotzky}, R.~F., \& {Szymkowiak}, A.~E. 1988, in NATO Advanced Science
  Institutes (ASI) Series C, Vol. 229, NATO Advanced Science Institutes (ASI)
  Series C, ed. A.~C. {Fabian}, 53

\bibitem[{{Norton} \& {Watson}(1989)}]{norton1989}
{Norton}, A.~J., \& {Watson}, M.~G. 1989, \mnras, 237, 853

\bibitem[{{Peterson} {et~al.}(2003){Peterson}, {Kahn}, {Paerels}, {Kaastra},
  {Tamura}, {Bleeker}, {Ferrigno}, \& {Jernigan}}]{peterson2003}
{Peterson}, J.~R., {Kahn}, S.~M., {Paerels}, F.~B.~S., {et~al.} 2003, \apj,
  590, 207

\bibitem[{{Porquet} \& {Dubau}(2000)}]{porquet2000}
{Porquet}, D., \& {Dubau}, J. 2000, \aaps, 143, 495

\bibitem[{{Porquet} {et~al.}(2001){Porquet}, {Mewe}, {Dubau}, {Raassen}, \&
  {Kaastra}}]{porquet2001}
{Porquet}, D., {Mewe}, R., {Dubau}, J., {Raassen}, A.~J.~J., \& {Kaastra},
  J.~S. 2001, \aap, 376, 1113

\bibitem[{{Reeves} {et~al.}(2008){Reeves}, {Done}, {Pounds}, {Terashima},
  {Hayashida}, {Anabuki}, {Uchino}, \& {Turner}}]{reeves2008}
{Reeves}, J., {Done}, C., {Pounds}, K., {et~al.} 2008, \mnras, 385, L108

\bibitem[{{Schlegel} {et~al.}(2014){Schlegel}, {Shipley}, {Rana}, {Barrett}, \&
  {Singh}}]{schlegel2014}
{Schlegel}, E.~M., {Shipley}, H.~V., {Rana}, V.~R., {Barrett}, P.~E., \&
  {Singh}, K.~P. 2014, \apj, 797, 38

\bibitem[{{Shaw} {et~al.}(2018){Shaw}, {Heinke}, {Mukai}, {Sivakoff},
  {Tomsick}, \& {Rana}}]{shaw2018}
{Shaw}, A.~W., {Heinke}, C.~O., {Mukai}, K., {et~al.} 2018, \mnras, 476, 554

\bibitem[{{Terada} {et~al.}(2004){Terada}, {Ishida}, \&
  {Makishima}}]{terada2004}
{Terada}, Y., {Ishida}, M., \& {Makishima}, K. 2004, \pasj, 56, 533

\bibitem[{{Williams} {et~al.}(1987){Williams}, {King}, \&
  {Brooker}}]{williams1987}
{Williams}, G.~A., {King}, A.~R., \& {Brooker}, J.~R.~E. 1987, \mnras, 226, 725

\end{thebibliography}
\end{document}